\documentclass[letterpaper,12pt]{article}
\usepackage{mathptmx}
\usepackage{graphics}
\usepackage{graphicx}
\usepackage[authoryear,comma,sectionbib]{natbib} 

\usepackage{amsmath}
\usepackage{amssymb}
\usepackage{color}

\begin{document}

\title{A clustering tool for nucleotide sequences using Laplacian Eigenmaps and Gaussian Mixture Models}

\author{
Marine BRUNEAU$^{(1)}$
\and
Thierry MOTTET$^{(2)}$
\and
Serge MOULIN$^{(2)}$
\thanks{
Corresponding author. Email {\tt serge.moulin@univ-fcomte.fr}
}
\and
Ma\"el KERBIRIOU$^{(1)}$
\and
Franz CHOULY$^{(1)}$
\and
St\'ephane CHRETIEN$^{(3)}$
\and 
Christophe GUYEUX$^{(2)}$
}

\date{\scriptsize
(1) Laboratoire de Math\'ematiques de Besan\c con, UMR 6623 CNRS, Universit\'e de Bourgogne Franche-Comt\'e, 16 route de Gray, 25030 Besan\c con, France.\\
(2) Computer Science Department, FEMTO-ST Institute, UMR 6174 CNRS, Universit\'e de Bourgogne Franche-Comt\'e, 15 Bis Avenue des Montboucons, 25030 Besan\c con, France.\\
(3) National Physical Laboratory, Hampton Road, Teddington, United Kingdom.
}





\maketitle

\begin{abstract}
We propose a new procedure for clustering nucleotide sequences based on the ``Laplacian Eigenmaps'' and Gaussian Mixture modelling. This proposal is then applied to a set of 100 DNA sequences from the mitochondrially encoded NADH dehydrogenase 3 (ND3) gene of a collection of \textit{Platyhelminthes} and \textit{Nematoda} species. The resulting clusters are then shown to be consistent with the gene phylogenetic tree computed using a maximum likelihood approach. This comparison shows in particular that the clustering produced by the methodology combining Laplacian Eigenmaps with Gaussian Mixture models is coherent with the phylogeny as well as with the NCBI taxonomy. We also developed a Python package for this procedure which is available online.

\end{abstract}

\section{Introduction}
\label{sec:Introduction}

As the amount of available genetic sequences increases drastically, the need for scalable methods becomes urgent. A very important tool for dealing with large amounts of data is clustering. Clustering was successfully used for a large number of important applications recently. 
Clustering methods have thus demonstrated irrefutable efficiency for real-life applications to genetics and have provided extremely powerful tools to the community.
In particular, the huge amount of 
molecular data available nowadays can help addressing new 
and essential questions in genomics if appropriate scalable methods are used. 
Sequence clustering is a key element among these tools~\citep{vgrm+15:ij}. Among the various benefits of sequence clustering are 
\begin{itemize}
   \item interpretability as for the classification of 16S ARN into OTU~\citep{hao2011clustering}, 
   \item potential reduction of very large databases, \textit{i.e.}, too large to be processed in a relatively short time~\citep{rousk2010soil}, or 
   \item creation of non-redundant protein groups~\citep{suzek2007uniref} 
    \end{itemize}
for instance. Moreover, clustering is also crucial for the discovery of new hidden variables underpinning the phenomena that are analysed. 

However, as noted by a number of researchers from the community of Machine Learning (ML), clustering may not be very reliable without an appropriate preliminary embedding of the data samples. The motto often quoted in ML is that high dimensional data are often too scattered for an off-the-shelf method to work properly.
Several embeddings have been proposed in the literature. Some of them are better suited for supervised learning, such as the methods based on neural networks (\textit{e.g.}, auto-encoders) and some other methods are more suitable for unsupervised learning as is the case in the present work. Among many nonlinear embedding methods, the Laplacian Eigenmap \citep{belkin2001laplacian}
approach has been extensively studied from both the theoretical side and the application one \citep{spielman2009spectral}. Such methods have received a lot of attention and can be also constrained towards better postprocessing with a view towards improved clustering as shown in, \textit{e.g.}, \citep{chretien2016semi}. 

In the case study, our goal is to show the practical efficiency of a combination of the plain Laplacian Eigenmap approach with Gaussian Mixture based clustering. Our choice of this combination is mainly motivated by scalability constraints and the fact that both methods are state of the art in the Machine Learning community.
The procedure is extremely natural and can be described as follows
\begin{enumerate}
\item Compute a similarity matrix between each couple of sequences. That is to say, a matrix $W$ of size $n \times n$ (where $n$ is the number of studied sequences), such that $W_{i,j}$ increases with the similarity of sequences $i$ and $j$.
\item Apply the Laplacian Eigenmap~\citep{belkin2001laplacian} method to the matrix $W$. This method transforms the sequences in elements leaving in a given vectorial space, and whose positions in the space reflects well and illustrates in an richer way the sequence similarities.
\item Apply a clustering method to the points of the yielding space using a Gaussian mixture model \citep{day1969estimating}.
\end{enumerate}
Our main contribution is twofold: we propose a ready to use Python package available online and we demonstrate the efficiency of the approach for the problem of clustering nucleotide sequences.

In particular, we tested the proposed methodology on a sample of one hundred ND3 genes (DNA sequences) from \textit{Platyhelminthes} and \textit{Nematoda} species that have been downloaded on the NCBI website\footnote{http://www.ncbi.nlm.nih.gov/}. The classification produced by the proposed method has been compared with the phylogenetic tree of these species obtained by a likelihood maximization method using PhyML~\citep{guindon2005phyml}. According to this comparison, the clustering is consistent with both clades appearing in the phylogenetic tree and the NCBI taxonomy.
More specifically, our clustering method perfectly separates the \textit{Nematoda} and \textit{Platyhelminthes} phyla. Among the \textit{Nematoda}, we obtained a cluster for \textit{Trichocephalida} order, another one for most of the species of the \textit{Spirurida} order, and a last cluster for the remainder. 

The plan of the paper is the following. The three stages of the proposed method are detailed in the next section, while an application example is provided in Section~\ref{sec:application}. The whole proposal is discussed in Section~\ref{Discussion}. Finally, this article ends with a conclusion section, in which the contribution is summarized and intended future work is outlined.

\section{The Clustering Method}

\subsection{Design of the similarity matrix}

The first step in the construction of a good embedding is the creation of a similarity matrix $W$. This matrix measures the similarity between every two sequences by providing a number between 0 and 1, the greater this number is, the most similar the sequences are. 

We introduce here a way to build similarity matrices using the Needleman Wunsch ``distance''~\citep{needleman1970general}. In order to do so, a multiple global alignment of the DNA sequences is run, for instance using MUSCLE (Multiple Sequence Comparison by Log-Expectation~\citep{edgar2004muscle}) software. Based on the results of this alignment, we can then define a first matrix $M$ satisfying the following requirements: for all $i,j \in [\![1,n]\!],$ $M_{i,j}$ is the "Needleman-Wunsch distance" between sequence $i$ and sequence $j$. More precisely, $M$ can be computed, by evaluating the Needleman-Wunsch edit distance between each couple of aligned sequences, which is the reference distance when considering nucleotidic sequences. In practice, this can be done using the \verb|needle| command from EMBOSS package~\citep{emboss}.

The matrix $M$ is then divided by the largest distance value, so that all its coefficients are between 0 and 1. $W$ can finally be obtained as follows: 
\begin{center}
$\forall$ $i,j \in [\![1,n]\!],$ $W_{i,j} = 1 - M_{i,j} ,$
\end{center}
in such a way that $W_{i,j}$ represents the similarity between species $ i $ and $ j $.

\subsection{Laplacian Eigenmaps on $W$}
\label{sec:laplacian}

The so-called Laplacian Eigenmaps~\citep{belkin2001laplacian} is an original method of representation, in a $k$-dimension vector space, of a given matrix of similarity between complex objects (here, words on the \{A,C,G,T\} alphabet). In this space, points with real coordinates are close when their similarity is large. Indeed, increasing the dimension leads to a richer view of the similarities between different objects, onto which standard clustering techniques can be applied.

The first step is to create the normalized Laplacian matrix~\citep{chen2007resistance}:
$$L = D^{-1/2}(D - W)D^{-1/2},$$ where $W$ is the similarity matrix defined previously and $D$ is the degree matrix of $W$. That is to say, $D$ is the diagonal matrix defined by:
\begin{center}$\forall i \in [\![1,n]\!],$ $D_{i,i} = \sum\limits_{j=1}^n W_{i,j}$ .\end{center}

\noindent $L$ being symmetric and real, it is diagonalisable in a basis of pairwise orthogonal eigenvectors $\left\{\phi_1, ..., \phi_n\right\}$ associated with eigenvalues $0=\lambda_1\leqslant\lambda_2\leqslant...\leqslant\lambda_n$.

The Laplacian Eigenmaps consists in considering the following embedding function:
$$c_{k_1}(i) = \left( \begin{array}{c}
\phi_2(i) \\
\phi_3(i) \\
\vdots \\
\phi_{k_1+1}(i) \\
\end{array} \right) \in \mathbb{R}^{k_1},$$
where $c_{k_1}(i)$ is the coordinate vector of the point 
corresponding to the $i^{th}$ sequence. In other words, the 
coordinate vector of the point corresponding to the 
$i^{th}$ sequence is constituted by the $i^{th}$ coordinate 
of each of the $k_1$ first eigenvectors, ordered according to the size of their eigenvalues. As stated previously, this cloud of points is a model of the DNA sequences, such that the proximity between two points increases (considering the Euclidean distance) with the sequence similarity.

The choice of the $k_1$ cutoff is a crucial issue and it is usually made as follows. The ordered eigenvalues are plotted, and we stop when the increase becomes negligible: the number of eigenvalues that are not discarded is $k_1$. For instance, in our program, we have chosen to set $k_1$ as the first time the difference between the $k^{th}$ and $(k+1)^{th}$ value is lower than 0.01.

Remark that $W$ can be seen as a weighted adjacency matrix of a graph, where nodes are the DNA sequences while edges are labeled by the degree of affinity between their adjacent nodes. In the literature, the Laplacian matrix is often described as constructed from the weighted adjacency matrix of such a graph rather than constructed from a similarity matrix. These definitions are equivalent.

\subsection{Gaussian Mixture based clustering}

The final step is performed by applying Gaussian Mixture based clustering  (GMM, \citep{day1969estimating}) to point cloud. Gaussian Mixture Models belong to the class of unsupervised learning schemes ~\citep{friedman2001elements}, and allows to distribute the data points into different clusters without \textit{a priori} assumption about the clusters' interpretation. One of the very useful features of model based clustering is that the model allows to use information criteria in order to estimate the number of clusters using AIC \citep{akaike1974new}, BIC \citep{schwarz1978estimating}, or ICL \citep{biernacki2000assessing}. The mathematical assumption of a GMM is that the point cloud follows the distribution:
$$\displaystyle \sum_{i=1}^{k_2} \delta_i \ \mathcal{N}(\mu_i, \Sigma_i),$$ 
where $k_2$ is the number of clusters, $\delta_i$ is the probability for a point to be in cluster $i$, and $\mathcal{N}(\mu_i, \Sigma_i)$ is the normal distribution of mean $\mu_i$ and covariance matrix $\Sigma_i$. GMM parameters are computed with the Expectation-Maximization (EM) algorithm~\citep{mclachlan2004finite}. Notice that 
the EM algorithm may converge to singular distributions exponentially fast \citep{biernacki2003degeneracy}. However, degenerate situations can be easily discarded and consistent estimators can be easly obtained in practice. Gaussian Mixture models are still a topic of current extensive research, both from the statistical perspective \citep{wang2014high} and the computational perspective \citep{yi2015regularized}.

We have chosen to consider the Bayesian Information Criterion (BIC,~\citep{schwarz1978estimating}) to determine the number of clusters $k_2$. The BIC, which is a criterion for model selection, is defined as follows:
$$BIC = -2 ln(L) + ln(n)p,$$ 
where $L$ is the likelihood of the estimated model, $n$ is the number of observations in the sample, and $p$ is the number of model parameters. 
This criterion allows us to select a model whose validity is based on a compromise between the value of the model likelihood (that we want to maximize) and the number of parameters to estimate (that we want to minimize). The likelihood of the model increases with $k_2$ as well as the number of parameters. The selected model will be by default the one that minimizes this criterion (but, in our software, the user can set manually, if needed, the desired number of clusters).

\subsection{The clustering software}
\label{sec:Algorithm}
The Python program corresponding to the algorithm described in this section is freely available online at https://github.com/SergeMOULIN/clustering-tool-for-nucleotide-sequences-using-Laplacian-Eigenmaps-and-Gaussian-Mixture-Models.
Its main function provides a clustering either from a matrix of distances or a similarity matrix. Its prototype meets the following canvas:
\begin{verbatim}
clustering = GClust(M, refs, simil = False,
       nbClusters = 'BIC', drawgraphs = True, 
       delta = 0.01)    
\end{verbatim}
where:
\begin{itemize}
\item[$\bullet$] \verb|M| is the matrix of distances between the genomic sequences. \verb|M| can also be a matrix of similarities between the genomic sequences, by setting 

\verb|simil = TRUE| (see below).
\item \verb|refs|, is the list of species references. Note that \verb|M| and \verb|refs| are the only mandatory arguments.
\item[$\bullet$] \verb|simil| is an optional Boolean argument. When \verb|simil = TRUE|, then \verb|M| is considered as a matrix of similarities between the DNA sequences. Otherwise it is a matrix of distances (\verb|simil = FALSE| by default).
\item[$\bullet$] \verb|nbClusters| is the number of clusters desired by the user. By default, the program applies the BIC criterium to determine it.
\item[$\bullet$] \verb|drawgraphs| is an optional Boolean value to produce some graphics. If \verb|drawgraphs = TRUE|, a two dimensional clustering of data is plotted (\textit{cf.} Figures~\ref{fig:graph1-2}, \ref{fig:graph1-3}, and \ref{fig:graph2-3}), as well as the graphical representation of similarities (as in Fig.~\ref{fig:graphsimil}).
\item[$\bullet$] \verb|delta| is the value used to select $k_1$ (\textit{i.e.}, the number of eigenvectors used in the Laplacian Eigenmaps). $k_1$ is the lowest value such that $\lambda_{k_1+1} - \lambda_{k_1} <$ \verb|delta|. \verb|delta| = 0.01 by default.
\end{itemize}

The output of \verb|Glust| function, named ``clustering'' above, is a list of $k_2$ lists. Each list contains references to the species grouped in a particular cluster.

\section{Illustration example on genomic data}
\label{sec:application}

\begin{figure}[!h]
\centering
\includegraphics[scale=0.35]{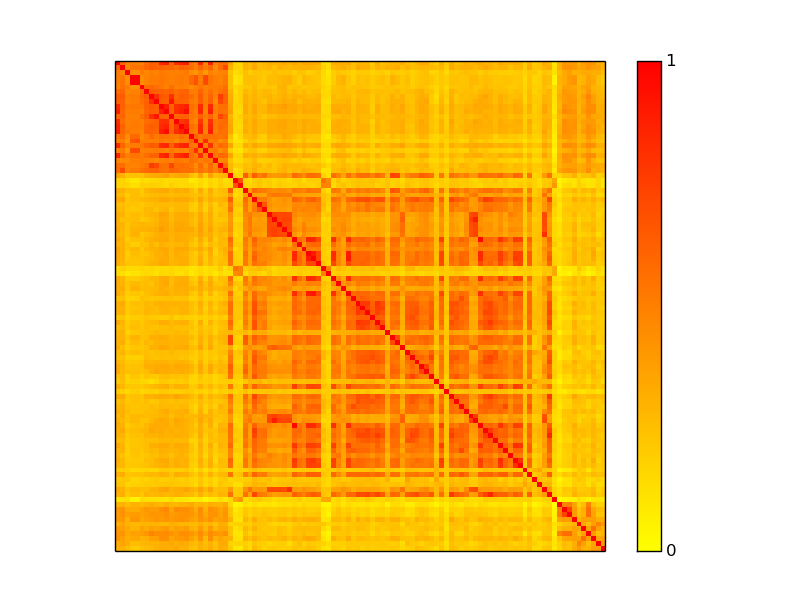}
\caption{Similarity matrix}
\label{fig:graphsimil}
\end{figure}

The proposed method has been applied to a sample of 100 DNA sequences from the mitochondrial gene ND3 taken from various species of both \textit{Platyhelminthes} and \textit{Nematoda}. 
Figure~\ref{fig:graphsimil} graphically displays the similarity matrix that has been obtained as described previously. As might be expected, the largest similarities are obtained on the diagonal, that is, when the sequences are compared to themselves. However, blocks seem to appear too in this representation, justifying the need to further investigate such similarities by applying our clustering method on these sequences.

\begin{figure}[!h]
\begin{center}
\includegraphics[scale=0.65]{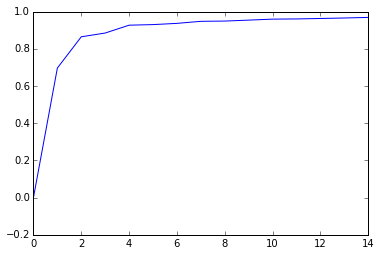}
\end{center}
\caption{Curve representing the first 14 eigenvalues}
\label{fig:eigenvalues}
\end{figure}

Figure~\ref{fig:eigenvalues} displays the 14 first ascending eigenvalues, labeled $\lambda_i$, and satisfying $0=\lambda_1\leq\lambda_2\leq...\leq\lambda_{14}$. According to the proposed methodology, we have chosen $k_1$ as the minimal index $k$ such that $\lambda_{k+1}-\lambda_{k}$ is lower than or equal to 0.01. In this case study, we found $k_1$ = 4.

\begin{figure}[!h]
\begin{center}
\includegraphics[scale=0.6]{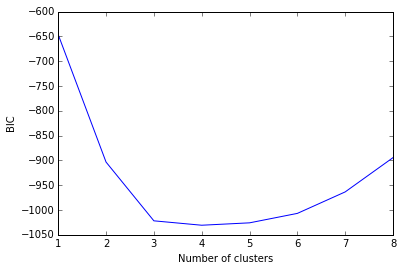}
\end{center}
\caption{Bayesian Information Criterion of the Gaussian Mixture Models}
\label{fig:BIC}
\end{figure}

Figure~\ref{fig:BIC} shows the Bayesian Information Criterion of the Gaussian Mixture Models for various number of clusters. One can see that this BIC reaches its minimum for $k_2 = 4$. Thus, the program automatically achieves the sequence partition in four clusters. 

\begin{figure}[!h]
\centering
\includegraphics[scale=0.43]{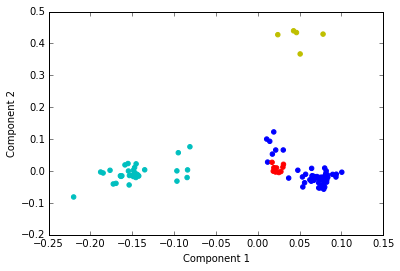}
\caption{GMM clustering in the plane formed by the eigenvectors 1 and 2}
\label{fig:graph1-2}
\end{figure}

\begin{figure}[!h]
\centering
\includegraphics[scale=0.43]{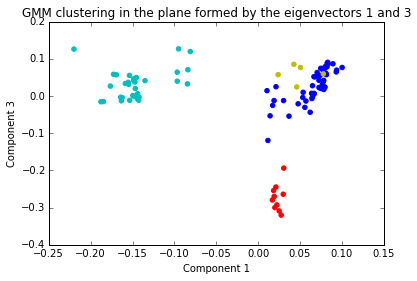}
\caption{GMM clustering in the plane formed by the eigenvectors 1 and 3}
\label{fig:graph1-3}
\end{figure}

\begin{figure}[!h]
\centering
\includegraphics[scale=0.43]{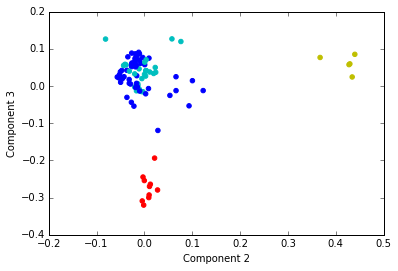}
\caption{GMM clustering in the plane formed by the eigenvectors 2 and 3}
\label{fig:graph2-3}
\end{figure}

Figures~\ref{fig:graph1-2},~\ref{fig:graph1-3}, 
and \ref{fig:graph2-3} represent the point cloud divided in 4 clusters.
This cloud point is projected into the plane formed by the first and second eigenvectors in 
Figure~\ref{fig:graph1-2}, to the one formed by the first and third eigenvectors in Figure~\ref{fig:graph1-3}, and to the plane formed by the second and the third eigenvectors in Figure~\ref{fig:graph2-3}.
In these graphs, clusters 0, 1, 2, and 3 are represented in red, cyan, blue, and yellow respectively.

\begin{figure*}[!h]
\centering
\includegraphics[scale=0.4]{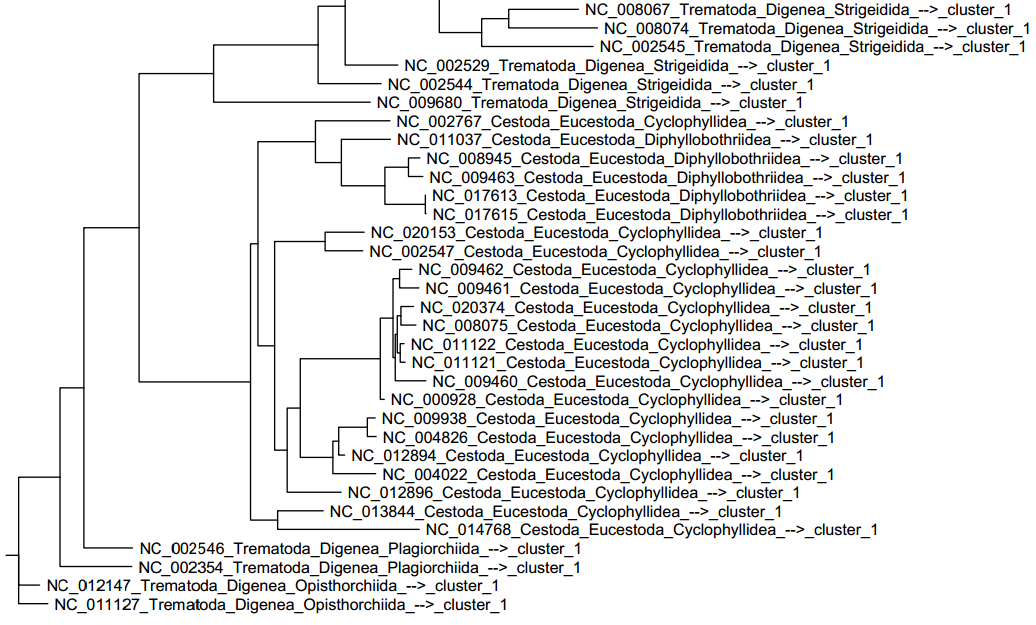}
\caption{First part of the phylogenetic tree (\textit{Platyhelminthes})}
\label{fig:Platyhelminthes}
\end{figure*}

\begin{figure*}[!h]
\centering
\includegraphics[scale=0.8]{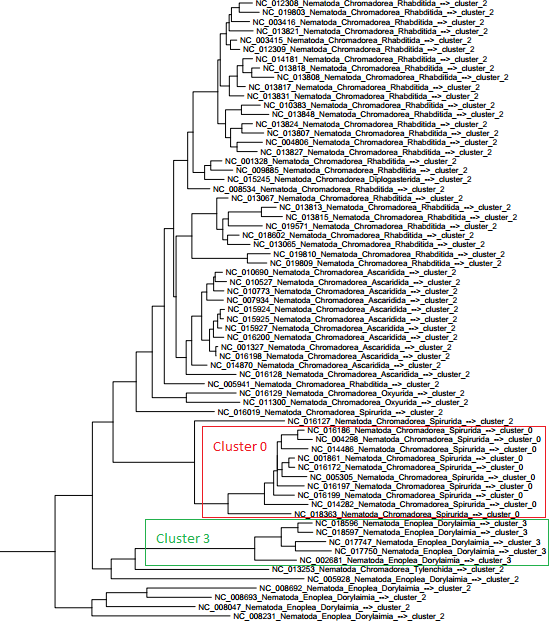}
\caption{Second part of the phylogenetic tree (\textit{Nematoda})}
\label{fig:Nematoda}
\end{figure*}

A FASTA file has then be written, in which each nucleotide sequence has been labeled according to its taxonomy and its cluster number. The taxonomies have been found thanks to the \verb|efetch| function (sub-package \verb|Entrez|\footnote{A package that provides code to access NCBI over the world wide web.}, Python package \verb|Bio|).
We then have used the online software \textit{PhyML} (maximum likelihood method method for phylogenetic tree reconstruction) with default options~\citep{guindon2005phyml}, to build a tree based on the same mitochondrial ND3 gene that we have used during clustering. The tree, whose leaves contain taxa names and cluster id, has been displayed using \textit{FigTree} software~\citep{morariu08figtree}. This tree is depicted in Figures~\ref{fig:Platyhelminthes} and~\ref{fig:Nematoda}.

The \verb|efetch| function provides 8 taxonomic levels for each species in our sample. In Figures~\ref{fig:Platyhelminthes} and~\ref{fig:Nematoda}, for readability reasons, we only show the taxonomy between levels 4 and 6.
Note first that this clustering perfectly separates \textit{Platyhelminthes} and \textit{Nematoda} phyla. Indeed \textit{Trematoda} and \textit{Cestoda} are two classes of \textit{Platyhelminthes}. 
Cluster 1 corresponds exactly to the \textit{Platyhelminthes}, while Clusters 0, 2, and 3 represent the \textit{Nematoda}. Cluster 0 is only composed of \textit{Spirurida}, it contains 10 of the 12 members of this taxon. More precisely, when considering the seventh taxonomic level, we found that Cluster 0 contains nine \textit{Filarioidea} and one \textit{Thelazioidea}, while the \textit{Spirurida} that are not in this cluster are a \textit{Dracunculoidea} and a \textit{Physalopteroidea}. Cluster 3, for its part, corresponds exactly with the \textit{Trichocephalida} taxon.

Finally, in addition to recover taxonomy, the clustering agrees very well with the tree obtained via PhyML, as shown in Figures~\ref{fig:Platyhelminthes} and \ref{fig:Nematoda}.
Note that the taxonomy, based on morphology and nuclear genome, fully agrees with the PhyML tree and our clustering, which are both based on the mitochondrial genome. 
This fact suggests that this mitochondrial gene evolves like the nuclear genome. 

\section{Discussion}
\label{Discussion}

Various options are possible to perform the analysis we have presented previously, some of them being listed hereafter.

\subsection{Similarity matrix}
\label{sec:Similarity matrix}
The multiple global alignment step can be achieved using other software than MUSCLE. Among the most extensively used methods, we could have chosen MAFFT~\citep{katoh2013mafft} for instance, as well as ClustalW or ClustalX \citep{larkin2007clustal}. It is also possible to apply a pairwise alignment method like the Needleman-Wunsch algorithm~\citep{needleman1970general}. Moreover, we have chosen EDNAFULL scoring matrix, but other matrices are available to produce a score between two aligned sequences, like PAM or BLOSUM. 
Finally, instead of defining the similarity matrix $W$ 
as $W_{i,j} = 1 - M_{i,j}$, it could be possible to consider $W_{i,j} = \frac{1}{M_{i,j}}$ or 
$W_{i,j} = e^{-M_{i,j}}$.

To find the best option among these possible choices, one can either test them on reference sequences on which the clusters to obtain are perfectly known, for instance by producing a laboratory-generated phylogeny. Another option can be to use a well-established phylogeny based on fossils morphology and molecular data. Finally, a third option can be to compute simulations of a set of species having a common ancestor. 
These three approaches have already been implemented to compare phylogenetic tree reconstruction tools. The first approach, for instance, has been applied in~\citep{hillis1992experimental} on the bacteriophage T7 evolved in a laboratory. The second approach, for its part, has been applied in~\citep{cummings1995sampling}, \citep{russo1996efficiencies}, and~\citep{zardoya1996phylogenetic}, who used mammalian or vertebrate phylogeny to perform these comparisons. The last one was considered a lot of time, and summarised in~\citep{felsenstein1988phylogenies}, \citep{huelsenbeck1995performance}, and~\citep{nei1996phylogenetic}.

\subsection{Number of considered eigenvectors}
\label{sec:Number of eigenvectors retained}

The number of eigenvectors to keep is another point to investigate.
As explained in Part~\ref{sec:laplacian}, it is usually 
advised to check graphically when the increase of eigenvalues is reducing. 
In this article, we have chosen to consider $k_1$ such that $\delta = \lambda_{k_1+1} - \lambda_{k_1} < 0.01$. 
This criterion has led to $k = 4$ in the case study, which seems acceptable according to the considered taxonomy. 

Some authors in the literature proposed to compute $k_1$ as the logarithm of $n$~\citep{matiasnotes}. It is also possible to consider a criterion related to the second derivative instead of the first one, by replacing the computation of $\delta$ by: 
$$\delta = (\lambda_{k_1} - \lambda_{k_1-1}) - (\lambda_{k_1+1} - \lambda_{k_1}).$$ 
The latter may be a good representative of the notion of inflection. Other approach can be considered to solve this problem, which is still an open one.

\subsection{Number of clusters}
\label{sec:Number of clusters}
We have chosen to consider the BIC~\citep{schwarz1978estimating} to determine the optimal number of clusters, which is a common choice for this type of problem. An alternative may be to use the Akaike Information Criterion (AIC,~\citep{akaike1974new}).
The principle of calculating the AIC is the same as the BIC, since the goal is to maximize log-likelihood penalized by the number of parameters (or more precisely, to minimize the number of parameters to which the log-likelihood is subtracted). AIC formula is the following:
$$ AIC = -2 ln(L) + 2 \times p,$$ where $L$ is the likelihood of the estimated model and $p$ the number of model parameters.

Indeed, BIC is ``more conservative'' than AIC. That is to say, the number of clusters obtained by BIC is lower or equal to the one obtained by AIC. In other words, BIC penalization is more important, and the choice between these two criteria can be dependent on how stringent the clustering is desired. 

\subsection{Conclusion}

In this work, we have proposed a new method of nucleotide sequence clustering. This clustering is produced by a methodology combining Laplacian Eigenmaps with Gaussian Mixture models, while the number of clusters is set according to the Bayesian Information Criterion. This methodology has been applied on
100 sequences of mitochondrially encoded NADH dehydrogenase 3. Obtained clusters are coherent with the phylogeny (gene tree obtained with PhyML) as well as with the NCBI taxonomy.

One possible extension for future work can be to investigate more deeply the impact of parameters in the obtained clusters. The effects of a modification in the similarity distance, in the manner to set the dimension of the space in Laplacian Eigenmaps, or in the number of desired clusters, will be systematically investigated. This clustering method will be applied in concrete problems related to genomics, for instance on coding sequences predicted on a set of bacteria, to determine the size of their core~\citep{acgmsb+14:ij} and pan genome.


\bibliographystyle{natbib}
\bibliography{biblio}

\end{document}